\documentclass[11pt,reqno]{amsart}

\usepackage[T1]{fontenc}
\usepackage[utf8]{inputenc}
\usepackage[english]{babel}

\usepackage[a4paper,margin=1.15in]{geometry}
\usepackage{microtype}
\usepackage{newpxtext}
\usepackage{mathtools}
\usepackage{amssymb}
\usepackage{newpxmath}
\usepackage{bm}
\usepackage{braket}
\allowdisplaybreaks
\numberwithin{equation}{section}

\usepackage{array}
\usepackage{tabularx}

\usepackage[shortlabels]{enumitem}
\usepackage[title]{appendix}
\usepackage[all,poly]{xy}
\usepackage{tikz}
\usetikzlibrary{arrows.meta,positioning}

\usepackage{xcolor}
\definecolor{linkblue}{RGB}{0,72,128}
\usepackage[
colorlinks=true,
linkcolor=linkblue,
citecolor=linkblue,
urlcolor=linkblue
]{hyperref}
\usepackage[nameinlink,noabbrev]{cleveref}

\theoremstyle{plain}
\newtheorem{thm}{Theorem}[section]
\newtheorem{theorem}[thm]{Theorem}

\newtheorem{corollary}[thm]{Corollary}
\newtheorem{corollary*}[thm]{Corollary*}

\newtheorem{proposition*}[thm]{Proposition*}

\theoremstyle{definition}

\newtheorem{defn-thm}[thm]{Definition-Theorem}

\theoremstyle{remark}
\newtheorem{remark}[thm]{Remark}

\newcommand{\Rmnum}[1]{\uppercase\expandafter{\romannumeral #1}}

\tikzset{
	solid black dot/.style={
		circle,
		fill=black,
		inner sep=1.5pt
	}
}

\def\d{\mathrm{d}}

\begin{document}
\raggedbottom
	
	 \title[Spectral curve for super Higher Weil-Petersson Volumes]{Eynard-Orantin Topological Recursions for Higher Weil-Petersson Volumes on Moduli Spaces of Super Riemann Surfaces}
	\author{Dingkang Xiong}
	\address{Department of Mathematical Sciences, Tsinghua University, Beijing 100084, China}
	\email{xdk22@mails.tsinghua.edu.cn}
	\author{Jian Zhou}
	\address{Department of Mathematical Sciences, Tsinghua University, Beijing 100084, China}
	\email{jianzhou@mail.tsinghua.edu.cn}

\begin{abstract}
		We derive an Eynard--Orantin topological recursion for intersection numbers involving Norbury's $\Theta$-class together with $\psi$-classes and higher $\kappa$-classes. These intersection numbers encode higher Weil--Petersson volumes of moduli spaces of super Riemann surfaces. The resulting recursion is a higher-$\kappa$ deformation of the topological recursion associated with the Bessel curve, which encodes the Br\'ezin--Gross--Witten model.
\end{abstract}
\maketitle

\section{Introduction}
	Two-dimensional topological gravity admits two closely related descriptions: matrix models and intersection theory on the Deligne--Mumford moduli spaces of stable curves. The Witten--Kontsevich theorem \cite{wittenTwodimensionalGravityIntersection1990,kontsevichIntersectionTheoryModuli1992} states that the partition function generating $\psi$-class intersection numbers is a distinguished tau-function of the KdV hierarchy, characterized by the string equation and the standard normalization. Equivalently, the same partition function satisfies the Virasoro constraints, which yield the Dijkgraaf--Verlinde--Verlinde (DVV) \cite{dijkgraafLoopEquationsVirasoro1991} recursion relations for the intersection numbers.
	
	More precisely, define the generating function (also called the free energy) of the intersection numbers on Deligne-Mumford moduli spaces by
	\begin{equation}\label{free energy}
		F(\mathbf{t};\hbar):=\sum_{\substack{g,n\geq 0\\2g-2+n>0}}\hbar^{2g-2}\sum_{a_1,\cdots,a_n\geq 0}\langle 	\tau_{a_1}\cdots\tau_{a_n}\rangle_g \frac{t_{a_1}\cdots t_{a_n}}{n!},
	\end{equation} 
	and define the partition function by
	\begin{equation}
		Z(\mathbf{t};\hbar):=\exp F(\mathbf{t};\hbar),
	\end{equation}	where 
	\begin{equation}\label{psi-class}
	\langle \tau_{a_1}\cdots\tau_{a_n}\rangle_g:=\int_{\overline{\mathcal{M}}_{g,n}}\psi^{a_1}_{1}\cdots \psi^{a_n}_{n},
	\end{equation}
	then the corresponding Virasoro constraints can be expressed as the following DVV recursion
	relations:
	\begin{equation}\label{DVV0}
		\begin{aligned}
			&(2a_{0}+1)!!\langle\tau_{a_0}\prod_{i=1}^{n}\tau_{a_i}\rangle_g= \\
			&\sum_{j=1}^{n}\frac{(2a_0+2a_j-1)!!}{(2a_j-1)!!}\langle\tau_{a_0+a_j-1}\prod_{i\ne 0,j}^{n}\tau_{a_i}\rangle_g \\
			&+\frac{1}{2}\sum_{b_1+b_2=a_0-2}(2b_1+1)!!(2b_2+1)!!\Bigg(\langle\tau_{b_1}\tau_{b_2}\prod_{i=1}^{n}\tau_{a_i}\rangle_{g-1} \\
			&+ \sum_{\substack{g_1+g_2=g\\[3pt]A_1\sqcup A_2=[n]}}^{\mathrm{stable}}\langle\tau_{b_1}\prod_{i\in A_1}\tau_{a_i}\rangle_{g_1}\cdot\langle\tau_{b_2}\prod_{i\in A_2}\tau_{a_i}\rangle_{g_2} \Bigg),
		\end{aligned}
	\end{equation}
	where $(-1)!!=1$, $[n]=\{1,2,\cdots,n\}$ and in the last term the summation $\sum^{\mathrm{stable}}$ is taken over the stable range: $2g_i-1+|A_i|>0$ for $i=1,2$.
	
	The DVV recursion relations have been generalized to many other intersection numbers on moduli spaces of stable curves. Motivated by Mirzakhani's celebrated recursion for the Weil--Petersson volumes of moduli spaces of bordered Riemann surfaces \cite{mirzakhaniSimpleGeodesicsWeilPetersson2006,mirzakhaniWeilPeterssonVolumesIntersection2006}, Mulase and Safnuk \cite{mulaseMirzakhanisRecursionRelations2006} generalized the DVV relations to intersection numbers containing the first $\kappa$-class $\kappa_1$. Furthermore, based on a series of works \cite{arbarelloCombinatorialAlgebrogeometricCohomology1994,faberAlgorithmsComputingIntersection1999,kaufmannHigherWeilPeterssonVolumes1996} about the relations between $\kappa$-classes and $\psi$-classes, Liu and Xu \cite{liuRecursionFormulaeHigher2008} generalized DVV relations to the intersection numbers corresponding to the higher Weil-Petersson volumes:
	\begin{equation}\label{h WP v}
		\langle \kappa_{b_1}\cdots\kappa_{b_m}\tau_{a_1}\cdots\tau_{a_n}\rangle_g
		:=\int_{\overline{\mathcal{M}}_{g,n}}\kappa_{b_1}\cdots\kappa_{b_m}\psi_{1}^{a_1}\cdots\psi_{n}^{a_n}.
	\end{equation}

	Weil--Petersson volumes also play a central role in Jackiw--Teitelboim (JT) gravity
 \cite{jackiwLowerDimensionalGravity1985,teitelboimGravitationHamiltonianStructure1983}.
Saad, Shenker, and Stanford
\cite{saadJTGravityMatrixIntegral2019}
identified the genus expansion of JT amplitudes with that of a double-scaled matrix integral, whose loop equations reproduce Mirzakhani's recursion in the Eynard--Orantin formulation.

The supersymmetric counterpart leads to the geometry of moduli spaces of super Riemann surfaces. Stanford and Witten \cite{stanfordJTGravityEnsembles2020}
related $\mathcal{N}=1$ JT supergravity to super
Weil--Petersson volumes and derived a supersymmetric analogue of Mirzakhani's recursion.
Norbury
\cite{norburyNewCohomologyClass2023,
norburyEnumerativeGeometryModuli2026}
constructed the classes $\Theta_{g,n}$ and expressed these
super volumes as intersection integrals on the ordinary moduli
space of stable curves:
\begin{equation}
  \widehat{V}_{g,n}(L_1,\ldots,L_n)
  =
  2^{1-g-n}
  \int_{\overline{\mathcal{M}}_{g,n}}
  \Theta_{g,n}
  \exp\left(
    2\pi^2\kappa_1
    +\frac12\sum_{i=1}^{n}L_i^2\psi_i
  \right).
\end{equation}
We use the normalization of
\cite{huangHigherWeilPeterssonVolumes2025}.

Under this identification, the Stanford--Witten recursion
becomes a recursion for intersection numbers involving
$\Theta_{g,n}$, $\psi$-classes, and powers of $\kappa_1$.
Huang, Liu, and Xu
\cite{huangHigherWeilPeterssonVolumes2025}
extended this recursion to arbitrary higher $\kappa$-classes.
In particular, they obtained the generalized DVV relations
\eqref{Recursion2} for the intersection numbers
\begin{equation}\label{super}
  \langle
    \kappa_{b_1}\cdots\kappa_{b_m}
    \tau_{a_1}\cdots\tau_{a_n}
  \rangle_g^\Theta
  :=
  \int_{\overline{\mathcal{M}}_{g,n}}
  \Theta_{g,n}
  \kappa_{b_1}\cdots\kappa_{b_m}
  \psi_1^{a_1}\cdots\psi_n^{a_n}.
\end{equation}


	
	Another poweful tool in the enumerative geometry is the Eynard--Orantin topological recursion theory \cite{eynardInvariantsAlgebraicCurves2007}. 
	Topological recursion provides a complementary framework for describing these intersection numbers. Given spectral-curve data 
	$$(\Sigma,x,y,B),$$ 
	where $\Sigma$ is a Riemann surface equipped with two suitable meromorphic functions $x$ and $y$, and $B$ is a symmetric bidifferential, usually called the Bergman kernel,
	the Eynard--Orantin recursion constructs a family of meromorphic multidifferentials $\omega_{g,n}$ on $\Sigma^n$. For appropriate spectral curves, these multidifferentials encode all the correlators.
	
	The spectral curve encoding the $\psi$-class intersection numbers \eqref{psi-class} is the Airy curve \cite{eynardInvariantsAlgebraicCurves2007,zhouTopologicalRecursionsEynard2013}:
	\begin{equation}
		x(z)=\frac{1}{2}z^2, \quad y(z)=z, \quad B(z_1,z_2)=\frac{\d z_1 \d z_2}{(z_1-z_2)^2},
	\end{equation}
and the  spectral curve encoding the higher Weil--Petersson volumes \eqref{h WP v}
has the same  $x$-function and the Bergman kernel $B$, whereas $y(z)=z$ is deformed to $y(z)=z+\sum_{k=2}^{\infty}u_k z^k$ (\cite{eynardIntersectionNumbersSpectral2011,eynardInvariantsSpectralCurves2014, eynardInvariantsAlgebraicCurves2007, zhouTopologicalRecursionsEynard2013}).	 
As shown by the  second-named author,
these topological recursions are equivalent to the DVV relations for the corresponding correlators.
	
	In \cite{doTopologicalRecursionBessel2018}, Do and Norbury showed that the partition function obtained by
	applying the Eynard--Orantin topological recursion to the Bessel curve
	\begin{equation}\label{Bessel curve}
		x(z)=\frac{1}{2}z^2,\qquad
		y(z)=\frac{1}{z},\qquad
		B(z_1,z_2)=\frac{\d z_1\,\d z_2}{(z_1-z_2)^2}
	\end{equation}
	coincides with the Br\'ezin--Gross--Witten (BGW) partition function \cite{brezinExternalFieldProblem1980,grossPossibleThirdorderPhase1980}.
	
	Norbury conjectured in \cite{norburyNewCohomologyClass2023} that the descendant potential of
	his $\Theta$-classes coincides with the BGW tau-function. This conjecture was
	proved by Chidambaram, Garc\'ia-Failde, and Giacchetto in
	\cite{chidambaramRelations$overlinemathcalM_gn$Negative2025}. Consequently, with the standard normalization,
	the BGW correlators are given by the intersection numbers
	\begin{equation*}
		\int_{\overline{\mathcal{M}}_{g,n}}
		\Theta_{g,n}\prod_{i=1}^{n}\psi_i^{d_i}.
	\end{equation*}

	In this article, following \cite{zhouTopologicalRecursionsEynard2013}, we use the generalized DVV relations of Huang, Liu, and Xu \cite{huangHigherWeilPeterssonVolumes2025} to derive a spectral curve encoding intersection numbers involving Norbury's $\Theta$-classes, higher $\kappa$-classes, and $\psi$-classes, as defined in \eqref{super}. The associated recursion is the Eynard--Orantin topological recursion for higher Weil--Petersson volumes on moduli spaces of super Riemann surfaces. The resulting spectral curve is a higher-$\kappa$ deformation of the Bessel curve.
	
	For a sequence with only finitely many terms nonzero $\mathbf{b}=(b(1),b(2),\ldots)\in\mathbb{N}^{\infty}$ and formal variables $\mathbf{s}=(s_1,s_2,\cdots)$, set 
	\[
	|\mathbf{b}|=\sum_{i\geq1}ib(i),\qquad
	\|\mathbf{b}\|=\sum_{i\geq1}b(i),\qquad
	\mathbf{b}!=\prod_{i\geq1}b(i)!,\qquad
	\mathbf{s}^{\mathbf{b}}=\prod_{i\geq1}s_i^{b(i)}.
	\]
	Then our main result in this article is
	
	\begin{theorem}\label{spectral curve for hsv}
		The Eynard--Orantin topological recursion for the spectral curve
		\begin{equation}
			\begin{aligned}
			x(z)=\frac{1}{2}z^2, \quad	y(z)=\sum_{\mathbf{b}\in\mathbb{N}^\infty}\frac{(-1)^{\Vert\mathbf{b}\Vert}\mathbf{s}^{\mathbf{b}}}{\mathbf{b}!(2|\mathbf{b}|-1)!!}z^{2|\mathbf{b}|-1}, \quad 
			B(z_1,z_2)=\frac{\d z_1 \d z_2}{(z_1-z_2)^2},
			\end{aligned}
		\end{equation}
		produces the multidifferentials
		\begin{equation}
				\omega^\Theta_{g,n}(z_1,\cdots,z_n;\mathbf{s}):=(-1)^n\sum_{a_1,\cdots,a_n\geq 0}\sum_{\ \mathbf{b}\in\mathbb{N}^\infty}\langle\kappa(\mathbf{b})\tau_{a_1}\cdots\tau_{a_n}\rangle_{g}^\Theta\cdot\frac{\mathbf{s}^\mathbf{b}}{\mathbf{b}!}\cdot\prod_{i=1}^n\frac{(2a_i+1)!!\d z_i}{z_i^{2a_i+2}},
		\end{equation}
		where
		\begin{equation}
			\langle\kappa(\mathbf{b})\tau_{a_1}\cdots\tau_{a_n}\rangle_{g}^\Theta=\int_{\overline{\mathcal{M}}_{g,n}}\Theta_{g,n}\prod_{i\geq 1}\kappa_{i}^{b(i)}\psi_{1}^{a_1}\cdots\psi_{n}^{a_n}.
		\end{equation}
	\end{theorem}
	\begin{remark}
		The deformed $y$-functions in the bosonic case, which corresponds to the higher Weil-Petersson volumes on moduli space of Riemann surfaces, and the super case are closely related. In the bosonic case, one has
		\begin{equation}\label{deform1}
			y^{\mathrm{bos}}(z)
			=
			\sum_{\mathbf{L}\in\mathbb{N}^{\infty}}
			\frac{(-1)^{\|\mathbf{L}\|}\mathbf{s}^{\mathbf{L}}}
			{\mathbf{L}!(2|\mathbf{L}|+1)!!}
			z^{2|\mathbf{L}|+1},
		\end{equation}
		whereas in the super case,
		\begin{equation}
			y^{\mathrm{super}}(z)
			=
			\sum_{\mathbf{L}\in\mathbb{N}^{\infty}}
			\frac{(-1)^{\|\mathbf{L}\|}\mathbf{s}^{\mathbf{L}}}
			{\mathbf{L}!(2|\mathbf{L}|-1)!!}
			z^{2|\mathbf{L}|-1}.
		\end{equation}
		Since $x(z)=z^2/2$, termwise differentiation gives
		\begin{equation}
			y^{\mathrm{super}}(z)
			=
			\frac{\d y^{\mathrm{bos}}(z)}{\d x(z)}.
		\end{equation}
		This relation was noted by Fuji and Manabe
		\cite{kobeuniversityjapanGeneralizationsMirzakhanisRecursion2024}.
	\end{remark}
	
	The Eynard-Orantin topological recursion defines the free energies $F_g$ for the spectral curve. We also show that, for genus $g\geq 2$, the free energy defined by the topological recursion coincides with the generating function of super higher Weil-Petersson volumes without $\psi$-classes.
	
	The rest of the paper is organized as follows. In Section $2$, we review the method in \cite{zhouTopologicalRecursionsEynard2013} that derives the Eynard-Orantin topological recursion from the DVV relation. In Section $3$, we derive the topological recursion and identify the corresponding
	spectral curve for super higher Weil-Petersson volumes. 

	\vspace{.1in}
{\bf Addition}.
This paper presents an early application of the approach of
\cite{zhouTopologicalRecursionsEynard2013,
zhouTopologicalRecursionVirasoro2024,
xiongEynardOrantinTopologicalRecursion2026}
to the setting of super moduli spaces.
We note that the same result has been proved in the recent
work of Ahmed and Naskar
\cite{ahmedDiscreteAnalogue$mathscrN2$2026}
using a different approach \cite{chekhovTopologicalRecursionHard2019}.

{\bf Acknowledgements}.
The authors thank Professor Kefeng Liu for communicating his work with his coauthors to us. 
The second-named author is partly supported by NSFC grants 11661131005
and 11890662.

{\bf Declaration of AI Assistance}.
No AI tools were used in developing the mathematical ideas, results, or proofs presented in this paper. AI tools were used solely to polish the
language of the manuscript. 
	
	\vskip 20pt
	
	\section{Eynard--Orantin Topological Recursion for \texorpdfstring{$n$}{n}-Point Functions on Moduli Spaces of Riemann Surfaces}
	In this section, we review the method of \cite{zhouTopologicalRecursionsEynard2013} for deriving the Eynard--Orantin topological recursion and its spectral curve from the DVV relations.
	
	In addition to the free energy defined in \eqref{free energy}, we introduce the $n$-point generating function
	\begin{equation}\label{npoint function}
		\begin{aligned}
			F_{g,n}(x_1,\ldots,x_n)
			&:=\int_{\overline{\mathcal{M}}_{g,n}}
			\exp\left(\sum_{i=1}^{n}x_i\psi_i\right)\\
			&=\sum_{a_1,\ldots,a_n\geq 0}
			\left\langle \tau_{a_1}\cdots\tau_{a_n}\right\rangle_g
			\prod_{i=1}^{n}\frac{x_i^{a_i}}{a_i!}.
		\end{aligned}
	\end{equation}
	Following \cite{eynardInvariantsAlgebraicCurves2007,eynardIntersectionNumbersSpectral2011,eynardInvariantsSpectralCurves2014}, we apply the following Laplace-type transform to $F_{g,n}(L_1^2,\ldots,L_n^2)$:
	\begin{equation}\label{recu func}
		\begin{aligned}
			W_{g,n}(z_1,\ldots,z_n)
			&:=\int_{\mathbb{R}_+^n}
			\left(\prod_{i=1}^{n}\d L_i\right)
			\exp\left(-\sqrt{2}\sum_{i=1}^{n}z_iL_i\right)
			\left(\prod_{i=1}^{n}2L_i\right)\\
			&\qquad\times
			F_{g,n}(L_1^2,\ldots,L_n^2)\\
			&=\sum_{a_1,\ldots,a_n\geq 0}
			\left\langle \tau_{a_1}\cdots\tau_{a_n}\right\rangle_g
			\prod_{i=1}^{n}\frac{(2a_i+1)!!}{z_i^{2a_i+2}}.
		\end{aligned}
	\end{equation}
	Thus, $F_{g,n}$ and $W_{g,n}$ determine one another coefficientwise. Substituting the DVV relation \eqref{DVV0} into \eqref{recu func} yields
	\begin{equation}\label{Wre}
		\begin{aligned}
			&W_{g,n+1}(z_0,z_1,\ldots,z_n)\\
			&=\sum_{a_0,\ldots,a_n\geq 0}\sum_{i=1}^{n}
			(2a_i+1)(2a_0+2a_i-1)!!
			\left\langle \tau_{a_0+a_i-1}
			\prod_{\substack{1\leq j\leq n\\j\neq i}}\tau_{a_j}\right\rangle_g\\
			&\qquad\qquad\times
			\frac{1}{z_0^{2a_0+2}z_i^{2a_i+2}}
			\prod_{\substack{1\leq j\leq n\\j\neq i}}
			\frac{(2a_j+1)!!}{z_j^{2a_j+2}}\\
			&\quad+\frac{1}{2}\sum_{a_0,\ldots,a_n\geq 0}
			\sum_{b_1+b_2=a_0-2}(2b_1+1)!!(2b_2+1)!!
			\frac{1}{z_0^{2a_0+2}}
			\prod_{i=1}^{n}\frac{(2a_i+1)!!}{z_i^{2a_i+2}}\\
			&\qquad\qquad\times\Bigg(
			\left\langle \tau_{b_1}\tau_{b_2}\prod_{i=1}^{n}\tau_{a_i}\right\rangle_{g-1}
			+\sum_{\substack{g_1+g_2=g\\A_1\sqcup A_2=[n]}}^{\mathrm{stable}}
			\left\langle \tau_{b_1}\prod_{i\in A_1}\tau_{a_i}\right\rangle_{g_1}
			\left\langle \tau_{b_2}\prod_{i\in A_2}\tau_{a_i}\right\rangle_{g_2}
			\Bigg).
		\end{aligned}
	\end{equation}
	The preceding relation admits the following residue formulation.
	\begin{theorem}[Zhou \cite{zhouTopologicalRecursionsEynard2013}, Theorem 2.2]\label{Airy re}
		The DVV recursion relation \eqref{DVV0} is equivalent to
		\begin{equation}\label{Wre cu}
			\begin{aligned}
				W_{g,n+1}(z_0,z_1,\ldots,z_n)
				&=\frac{1}{2}\operatorname*{Res}_{z=0}K(z_0,z)
				\Bigg[W_{g-1,n+2}(z,-z,z_{[n]})\\
				&\qquad+\sum_{\substack{g_1+g_2=g\\A_1\sqcup A_2=[n]}}
				W_{g_1,|A_1|+1}(z,z_{A_1})
				W_{g_2,|A_2|+1}(-z,z_{A_2})\Bigg],
			\end{aligned}
		\end{equation}
		where
		\begin{equation}\label{reker0}
			K(z,w):=\sum_{l=0}^{\infty}\frac{w^{2l-1}}{z^{2l+2}}
			=\frac{1}{w(z^2-w^2)},
		\end{equation}
		with initial values
		\begin{equation}
			W_{0,1}:=0,\qquad W_{0,2}(z,w):=\frac{1}{(z-w)^2}.
		\end{equation}
		Here $z_{[n]}:=(z_1,\ldots,z_n)$ and $z_A:=(z_i)_{i\in A}$ for $A\subseteq[n]$.
	\end{theorem}
	
	To identify \eqref{Wre cu} with topological recursion, recall the local Eynard--Orantin formalism for a spectral curve with a unique simple branch point at $z=0$. The relevant spectral-curve data are
	\begin{equation}\label{spectral curve}
		x(z)=\frac{1}{2}z^2,\qquad
		\omega_{0,1}(z)=y(z)\d x(z),\qquad
		\omega_{0,2}(z_1,z_2)=B(z_1,z_2),
	\end{equation}
	where $y$ is meromorphic near $z=0$. The recursion kernel is
	\begin{equation}\label{recursion kernel}
		\mathbb{K}(z_0,z)
		:=\frac{\displaystyle\int_{z'=-z}^{z}B(z_0,z')}
		{\bigl(y(z)-y(-z)\bigr)\d x(z)}.
	\end{equation}
	For the Airy curve, $y(z)=z$ and
	\begin{equation}\label{Bergman kernel}
		B(z_1,z_2)=\frac{\d z_1\,\d z_2}{(z_1-z_2)^2}.
	\end{equation}
	A direct calculation then gives
	\begin{equation}
		\mathbb{K}(z_0,z)=\frac{\d z_0}{\d z}\,
		\frac{1}{z(z_0^2-z^2)}.
	\end{equation}
	
	For stable $(g,n+1)$, the topological recursion produces the multidifferentials $\omega_{g,n+1}$ according to
	\begin{equation}
		\begin{aligned}
			\omega_{g,n+1}(z_0,z_1,\ldots,z_n)
			&=\frac{1}{2}\operatorname*{Res}_{z=0}\mathbb{K}(z_0,z)
			\Bigg[\omega_{g-1,n+2}(z,-z,z_{[n]})\\
			&\qquad+\sum_{\substack{g_1+g_2=g\\A_1\sqcup A_2=[n]}}^{\mathrm{stable}}
			\omega_{g_1,|A_1|+1}(z,z_{A_1})
			\omega_{g_2,|A_2|+1}(-z,z_{A_2})\Bigg].
		\end{aligned}
	\end{equation}
	Here the superscript ``stable'' indicates that terms involving $\omega_{0,1}$ are omitted.
	
	Comparing this recursion with Theorem~\ref{Airy re}, we obtain
	\begin{equation}
		\mathbb{K}(z,w)=K(z,w)\frac{\d z}{\d w},\qquad
		\omega_{0,2}(z_1,z_2)=W_{0,2}(z_1,z_2)\d z_1\,\d z_2.
	\end{equation}
	An induction on $2g-2+n$ now yields
	\begin{equation}
		\omega_{g,n}(z_1,\ldots,z_n)
		=W_{g,n}(z_1,\ldots,z_n)\prod_{i=1}^{n}\d(-z_i).
	\end{equation}
	Consequently, the Eynard--Orantin topological recursion on the Airy curve reproduces the $\psi$-class intersection numbers on the moduli spaces of stable curves.
	
\vskip 20pt

\section{Spectral Curve for Higher Weil--Petersson Volumes on Moduli Spaces of Super Riemann Surfaces}
In this section, we derive an Eynard--Orantin topological recursion and identify the corresponding spectral curve for higher Weil--Petersson volumes on moduli spaces of super Riemann surfaces.

In analogy with \eqref{recu func}, introduce formal parameters $\mathbf{s}=(s_1,s_2,\ldots)$ and define the generating functions
\begin{equation}\label{WTheta definition}
	W^\Theta_{g,n}(z_1,\ldots,z_n;\mathbf{s})
	:=\sum_{a_1,\ldots,a_n\geq 0}
	\sum_{\mathbf{b}\in\mathbb{N}^\infty}
	\left\langle\kappa(\mathbf{b})\tau_{a_1}\cdots\tau_{a_n}\right\rangle_{g}^\Theta
	\frac{\mathbf{s}^{\mathbf{b}}}{\mathbf{b}!}
	\prod_{i=1}^n\frac{(2a_i+1)!!}{z_i^{2a_i+2}}.
\end{equation}
Here $\mathbb{N}^\infty$ denotes the set of sequences $\mathbf{m}=(m(1),m(2),\ldots)$ of nonnegative integers with finite support. We use the notation
\begin{equation*}
	|\mathbf{m}|:=\sum_{i\geq 1}im(i),\qquad
	\|\mathbf{m}\|:=\sum_{i\geq 1}m(i),\qquad
	\mathbf{s}^{\mathbf{m}}:=\prod_{i\geq 1}s_i^{m(i)},\qquad
	\mathbf{m}!:=\prod_{i\geq 1}m(i)!,
\end{equation*}
and, whenever the entries are defined,
\begin{equation*}
	\binom{\mathbf{m}}{\mathbf{t}}
	:=\prod_{i\geq 1}\binom{m(i)}{t(i)},\qquad
	\binom{\mathbf{m}}{\mathbf{a}_1,\ldots,\mathbf{a}_n}
	:=\prod_{i\geq 1}\binom{m(i)}{a_1(i),\ldots,a_n(i)}.
\end{equation*}
For $\mathbf{b}\in\mathbb{N}^\infty$, set
\begin{equation}
	\kappa(\mathbf{b}):=\prod_{i\geq 1}\kappa_i^{b(i)}.
\end{equation}
As in \eqref{super}, the corresponding intersection numbers are
\begin{equation}
	\left\langle\kappa(\mathbf{b})\tau_{a_1}\cdots\tau_{a_n}\right\rangle_{g}^\Theta
	:=\int_{\overline{\mathcal{M}}_{g,n}}
	\Theta_{g,n}\prod_{i\geq 1}\kappa_i^{b(i)}
	\prod_{j=1}^n\psi_j^{a_j}.
\end{equation}
Such an intersection number can be nonzero only if
\begin{equation}
	|\mathbf{b}|+\sum_{i=1}^{n}a_i
	=\sum_{i\geq 1}ib(i)+\sum_{i=1}^{n}a_i
	=g-1.
\end{equation}
In genus zero, all these intersection numbers vanish. In genus one, the only potentially nonzero terms are $\langle\tau_0^n\rangle_1^\Theta$. Thus, nontrivial $\kappa$-class insertions can occur only when $g\geq 2$.

We begin with the generalized DVV recursion of Huang, Liu, and Xu.
\begin{theorem}[Huang--Liu--Xu, Theorem~1.4 \cite{huangHigherWeilPeterssonVolumes2025}]\label{RecursionB}
	Let $\mathbf{b}\in\mathbb{N}^{\infty}$, $g\geq 1$, $n\geq 0$, and $a_j\geq 0$. Then
	\begin{equation}\label{Recursion2}
		\begin{aligned}
			&(2a_0+1)!!
			\left\langle\kappa(\mathbf{b})\tau_{a_0}\prod_{i=1}^{n}\tau_{a_i}\right\rangle^{\Theta}_g\\
			&=\sum_{j=1}^{n}\sum_{\mathbf{L}+\mathbf{L}'=\mathbf{b}}
			\alpha_{\mathbf{L}}\binom{\mathbf{b}}{\mathbf{L}}
			\frac{(2|\mathbf{L}|+2a_0+2a_j+1)!!}{(2a_j-1)!!}
			\left\langle\kappa(\mathbf{L}')\tau_{|\mathbf{L}|+a_0+a_j}
			\prod_{\substack{1\leq i\leq n\\i\neq j}}\tau_{a_i}\right\rangle^{\Theta}_g\\
			&\quad+\frac{1}{2}\sum_{\mathbf{L}+\mathbf{L}'=\mathbf{b}}
			\sum_{r+s=|\mathbf{L}|+a_0-1}
			\alpha_{\mathbf{L}}\binom{\mathbf{b}}{\mathbf{L}}(2r+1)!!(2s+1)!!
			\left\langle\kappa(\mathbf{L}')\tau_r\tau_s
			\prod_{i=1}^{n}\tau_{a_i}\right\rangle^{\Theta}_{g-1}\\
			&\quad+\frac{1}{2}\sum_{\mathbf{L}+\mathbf{e}+\mathbf{f}=\mathbf{b}}
			\sum_{r+s=|\mathbf{L}|+a_0-1}
			\sum^{\mathrm{stable}}_{\substack{g_1+g_2=g\\[2pt]I\sqcup J=[n]}}
			\alpha_{\mathbf{L}}\binom{\mathbf{b}}{\mathbf{L},\mathbf{e},\mathbf{f}}
			(2r+1)!!(2s+1)!!\\
			&\qquad\times
			\left\langle\kappa(\mathbf{e})\tau_r\prod_{i\in I}\tau_{a_i}\right\rangle^{\Theta}_{g_1}
			\left\langle\kappa(\mathbf{f})\tau_s\prod_{j\in J}\tau_{a_j}\right\rangle^{\Theta}_{g_2}.
		\end{aligned}
	\end{equation}
	The coefficients $\alpha_{\mathbf{L}}$ are determined by
	\begin{equation}
		\sum_{\mathbf{L}+\mathbf{L}'=\mathbf{b}}
		\frac{(-1)^{\|\mathbf{L}\|}\alpha_{\mathbf{L}}}
		{\mathbf{L}!\mathbf{L}'!(2|\mathbf{L}'|-1)!!}=0,
		\qquad \mathbf{b}\neq\mathbf{0},
	\end{equation}
	or equivalently,
	\begin{equation}
		\alpha_{\mathbf{b}}
		=\mathbf{b}!\sum_{\substack{\mathbf{L}+\mathbf{L}'=\mathbf{b}\\[1pt]\mathbf{L}'\neq\mathbf{0}}}
		\frac{(-1)^{\|\mathbf{L}'\|-1}\alpha_{\mathbf{L}}}
		{\mathbf{L}!\mathbf{L}'!(2|\mathbf{L}'|-1)!!},
		\qquad \mathbf{b}\neq\mathbf{0},
	\end{equation} 
	with initial value $\alpha_{\mathbf{0}}=1$.
\end{theorem}

Define
\begin{equation}
	\beta_{\mathbf{b}}:=\frac{\alpha_{\mathbf{b}}}{\mathbf{b}!}.
\end{equation}
The defining identity for $\alpha_{\mathbf{b}}$ may then be written as
\begin{equation}
	\sum_{\mathbf{L}+\mathbf{L}'=\mathbf{b}}
	\frac{(-1)^{\|\mathbf{L}'\|}\beta_{\mathbf{L}}}
	{\mathbf{L}'!(2|\mathbf{L}'|-1)!!}=0,
	\qquad \mathbf{b}\neq\mathbf{0}.
\end{equation}
Multiplying by $\mathbf{s}^{\mathbf{b}}$ and summing over all $\mathbf{b}$ gives
\begin{equation}
	\left(\sum_{\mathbf{L}\in\mathbb{N}^\infty}
	\beta_{\mathbf{L}}\mathbf{s}^{\mathbf{L}}\right)
	\left(\sum_{\mathbf{L}'\in\mathbb{N}^\infty}
	\frac{(-1)^{\|\mathbf{L}'\|}}{(2|\mathbf{L}'|-1)!!}
	\frac{\mathbf{s}^{\mathbf{L}'}}{\mathbf{L}'!}\right)=1,
\end{equation}
where the constant term on the right-hand side corresponds to $\mathbf{b}=\mathbf{0}$. Consequently,

\begin{equation}\label{beta inverse}
	\sum_{\mathbf{L}\in\mathbb{N}^\infty}\beta_{\mathbf{L}}\mathbf{s}^{\mathbf{L}}
	=\left(
	\sum_{\mathbf{L}'\in\mathbb{N}^\infty}
	\frac{(-1)^{\|\mathbf{L}'\|}}{(2|\mathbf{L}'|-1)!!}
	\frac{\mathbf{s}^{\mathbf{L}'}}{\mathbf{L}'!}
	\right)^{-1}.
\end{equation}

The recursion starts from the initial values
\begin{equation}
	\left\langle\tau_0\right\rangle_1^\Theta=\frac{1}{8},
	\qquad
	\left\langle\kappa_1\right\rangle_2^\Theta=\frac{3}{128}.
\end{equation}
The first few correlators obtained from the generalized DVV relation are
\begin{equation}
	\left\langle\cdots\right\rangle_0^\Theta=0,
\end{equation}
\begin{equation}
	\left\langle\tau_0^n\right\rangle_1^\Theta=\frac{(n-1)!}{8},
\end{equation}
\begin{equation}
	\left\langle\tau_1\right\rangle_2^\Theta=\frac{3}{128},\qquad
	\left\langle\kappa_1\tau_0\right\rangle_2^\Theta=\frac{9}{128},\qquad
	\left\langle\tau_0\tau_1\right\rangle_2^\Theta=\frac{9}{128},\qquad
	\left\langle\kappa_1\tau_0^2\right\rangle_2^\Theta=\frac{9}{32}.
\end{equation}
Accordingly, the first few generating functions are
\begin{equation}
	W_{0,n}^\Theta=0\qquad\text{for }n\geq 3,
\end{equation}
\begin{equation}\label{W1n}
	W^{\Theta}_{1,n}(z_1,\ldots,z_n;\mathbf{s})
	=\frac{(n-1)!}{8z_1^2\cdots z_n^2},
\end{equation}
\begin{equation}
	W^\Theta_{2,1}(z_1;\mathbf{s})
	=\frac{9}{128z_1^4}+\frac{9s_1}{128z_1^2},
\end{equation}
\begin{equation}
	W^\Theta_{2,2}(z_1,z_2;\mathbf{s})
	=\frac{27}{128z_1^4z_2^2}
	+\frac{27}{128z_1^2z_2^4}
	+\frac{9s_1}{32z_1^2z_2^2}.
\end{equation}

As in the derivation of \eqref{Wre}, substituting the generalized DVV relation into \eqref{WTheta definition} gives the following recursion for $W^\Theta_{g,n}$:
\begin{equation}\label{Wtheta}
	\begin{aligned}
		&W^\Theta_{g,n+1}(z_0,z_1,\ldots,z_n;\mathbf{s})\\
		&=\sum_{a_0,a_1,\ldots,a_n\geq 0}
		\sum_{\mathbf{L}\in\mathbb{N}^\infty}\beta_\mathbf{L}\mathbf{s}^\mathbf{L}
		\Bigg[
		\sum_{i=1}^{n}(2a_i+1)(2|\mathbf{L}|+2a_0+2a_i+1)!! \\
		&\qquad\times\sum_{\mathbf{L}'\in\mathbb{N}^\infty}
		\left\langle\kappa(\mathbf{L}')\tau_{|\mathbf{L}|+a_0+a_i}
		\prod_{\substack{1\leq j\leq n\\j\neq i}}\tau_{a_j}\right\rangle_{g}^\Theta
		\frac{\mathbf{s}^{\mathbf{L}'}}{\mathbf{L}'!}
		\frac{1}{z_0^{2a_0+2}z_i^{2a_i+2}}
		\prod_{\substack{1\leq j\leq n\\j\neq i}}\frac{(2a_j+1)!!}{z_j^{2a_j+2}}\\
		&\quad+\frac{1}{2}\sum_{b_1+b_2=|\mathbf{L}|+a_0-1}
		(2b_1+1)!!(2b_2+1)!!
		\frac{1}{z_0^{2a_0+2}}
		\prod_{i=1}^{n}\frac{(2a_i+1)!!}{z_i^{2a_i+2}}\\
		&\qquad\times\Bigg(
		\sum_{\mathbf{L}'\in\mathbb{N}^\infty}
		\left\langle\kappa(\mathbf{L}')\tau_{b_1}\tau_{b_2}
		\prod_{i=1}^n\tau_{a_i}\right\rangle_{g-1}^\Theta
		\frac{\mathbf{s}^{\mathbf{L}'}}{\mathbf{L}'!}\\
		&\qquad\quad+\sum^{\mathrm{stable}}_{\substack{g_1+g_2=g\\[2pt]A_1\sqcup A_2=[n]}}
		\left(
		\sum_{\mathbf{e}\in\mathbb{N}^\infty}
		\left\langle\kappa(\mathbf{e})\tau_{b_1}\prod_{i\in A_1}\tau_{a_i}\right\rangle_{g_1}^\Theta
		\frac{\mathbf{s}^\mathbf{e}}{\mathbf{e}!}\right)\\
		&\hspace{195pt}\times
		\left(
		\sum_{\mathbf{f}\in\mathbb{N}^\infty}
		\left\langle\kappa(\mathbf{f})\tau_{b_2}\prod_{i\in A_2}\tau_{a_i}\right\rangle_{g_2}^\Theta
		\frac{\mathbf{s}^\mathbf{f}}{\mathbf{f}!}\right)
		\Bigg)		
		\Bigg].
	\end{aligned}
\end{equation}

This coefficient recursion admits the following residue formulation.
\begin{theorem}\label{Theta residue recursion}
	The generalized DVV relation \eqref{Recursion2} is equivalent to
	\begin{equation}\label{WTheta}
		\begin{aligned}
			&W_{g,n+1}^\Theta(z_0,z_1,\ldots,z_n;\mathbf{s})\\
			&=\frac{1}{2}\operatorname*{Res}_{z=0}K(z_0,z;\mathbf{s})
			\Bigg[W_{g-1,n+2}^\Theta(z,-z,z_{[n]};\mathbf{s})\\
			&\qquad+\sum_{\substack{g_1+g_2=g\\[2pt]A_1\sqcup A_2=[n]}}
			W_{g_1,|A_1|+1}^\Theta(z,z_{A_1};\mathbf{s})
			W_{g_2,|A_2|+1}^\Theta(-z,z_{A_2};\mathbf{s})\Bigg],
		\end{aligned}
	\end{equation}
	where
	\begin{equation}
		\begin{aligned}
			K(z,w;\mathbf{s})
			&:=\sum_{\mathbf{L}\in\mathbb{N}^\infty}
			\beta_{\mathbf{L}}\mathbf{s}^{\mathbf{L}}
			\sum_{l\geq|\mathbf{L}|}\frac{w^{2l+1}}{z^{2l+2-2|\mathbf{L}|}}\\
			&=\frac{w}{z^2-w^2}
			\sum_{\mathbf{L}\in\mathbb{N}^\infty}
			w^{2|\mathbf{L}|}\beta_{\mathbf{L}}\mathbf{s}^{\mathbf{L}},
		\end{aligned}
	\end{equation}
	with initial values
	\begin{equation}
		W_{0,1}^\Theta:=0,
		\qquad
		W_{0,2}^\Theta(z,w;\mathbf{s}):=\frac{1}{(z-w)^2}.
	\end{equation}
	As before, $z_{[n]}=(z_1,\ldots,z_n)$ and $z_A=(z_i)_{i\in A}$.
\end{theorem}
\begin{proof}
	The result follows by a direct residue calculation. In the quadratic term of \eqref{WTheta}, the contributions with
	\begin{equation*}
		(g_1,A_1)=(0,\{i\}),\qquad (g_2,A_2)=(g,[n]\setminus\{i\}),
	\end{equation*}
	and the analogous terms with the two factors interchanged reproduce the first term on the right-hand side of \eqref{Wtheta}. The genus-reduction contribution and the remaining stable splittings reproduce the two terms inside the second large parenthesis in \eqref{Wtheta}.
\end{proof}
\begin{remark}
	For $(g,n)\neq(0,2)$, the function $W^\Theta_{g,n}$ is even in each variable. It is therefore convenient to replace $W^\Theta_{0,2}$ in the recursion by its even part,
	\begin{equation}
		\widehat{W}^\Theta_{0,2}(z_1,z_2;\mathbf{s})
		:=\frac{z_1^2+z_2^2}{(z_1^2-z_2^2)^2},
	\end{equation}
	as in \cite{zhouTopologicalRecursionsEynard2013}. This replacement does not affect any of the residues defining the stable correlators.
\end{remark}

We illustrate \eqref{WTheta} with two elementary computations. First,
\begin{equation*}
	\begin{aligned}
		&W^\Theta_{0,3}(z_0,z_1,z_2;\mathbf{s})\\
		&=\operatorname*{Res}_{z=0}K(z_0,z;\mathbf{s})
		\widehat{W}_{0,2}^\Theta(z,z_1;\mathbf{s})
		\widehat{W}_{0,2}^\Theta(z,z_2;\mathbf{s})\\
		&=\operatorname*{Res}_{z=0}
		\Bigg(\sum_{\mathbf{L}\in\mathbb{N}^\infty}
		\beta_\mathbf{L}\mathbf{s}^\mathbf{L}
		\sum_{l\geq |\mathbf{L}|}
		\frac{z^{2l+1}}{z_0^{2l+2-2|\mathbf{L}|}}
		\frac{z^2+z_1^2}{(z^2-z_1^2)^2}
		\frac{z^2+z_2^2}{(z^2-z_2^2)^2}\Bigg).
	\end{aligned}
\end{equation*}
The integrand is holomorphic at $z=0$, and hence
\begin{equation*}
	W^\Theta_{0,3}(z_0,z_1,z_2;\mathbf{s})=0.
\end{equation*}
An immediate induction gives
\begin{equation*}
	W^\Theta_{0,n}(z_1,\ldots,z_n;\mathbf{s})=0
	\qquad\text{for }n\geq 3.
\end{equation*}
For genus one, the case $n=1$ gives
\begin{equation*}
	\begin{aligned}
		&W^\Theta_{1,1}(z_0;\mathbf{s})\\
		&=\frac{1}{2}\operatorname*{Res}_{z=0}
		K(z_0,z;\mathbf{s})W^\Theta_{0,2}(z,-z;\mathbf{s})\\
		&=\frac{1}{2}\operatorname*{Res}_{z=0}
		\Bigg(\sum_{\mathbf{L}\in\mathbb{N}^\infty}
		\beta_\mathbf{L}\mathbf{s}^\mathbf{L}
		\sum_{l\geq |\mathbf{L}|}
		\frac{z^{2l+1}}{z_0^{2l+2-2|\mathbf{L}|}}
		\frac{1}{(2z)^2}\Bigg)\\
		&=\frac{1}{8z_0^2}.
	\end{aligned}
\end{equation*}
More generally, for $n\geq 1$, induction on $n$ yields
\begin{equation*}
	\begin{aligned}
		&W^\Theta_{1,n+1}(z_0,z_1,\ldots,z_n;\mathbf{s})\\
		&=\frac{1}{2}\operatorname*{Res}_{z=0}K(z_0,z;\mathbf{s})
		\sum_{i=1}^{n}\Bigg(
		W^\Theta_{0,2}(z,z_i;\mathbf{s})
		W^\Theta_{1,n}(-z,z_1,\ldots,\widehat{z_i},\ldots,z_n)\\
		&\hspace{135pt}+
		W^\Theta_{0,2}(-z,z_i;\mathbf{s})
		W^\Theta_{1,n}(z,z_1,\ldots,\widehat{z_i},\ldots,z_n)
		\Bigg)\\
		&=\operatorname*{Res}_{z=0}
		\Bigg(\sum_{\mathbf{L}\in\mathbb{N}^\infty}
		\beta_\mathbf{L}\mathbf{s}^\mathbf{L}
		\sum_{l\geq |\mathbf{L}|}
		\frac{z^{2l+1}}{z_0^{2l+2-2|\mathbf{L}|}}\\
		&\qquad\times\sum_{i=1}^{n}
		\frac{z^2+z_i^2}{(z^2-z_i^2)^2}
		\frac{(n-1)!}{8z^2z_1^2\cdots\widehat{z_i^2}\cdots z_n^2}
		\Bigg)\\
		&=\frac{n!}{8z_0^2\cdots z_n^2}.
	\end{aligned}
\end{equation*}
This agrees with \eqref{W1n}.

We have thus recast the generalized DVV relation of Huang, Liu, and Xu \cite{huangHigherWeilPeterssonVolumes2025} as a residue recursion of Eynard--Orantin type. We now compare it with the topological recursion associated with the spectral-curve data in \eqref{spectral curve}. Since only the odd part of $y$ contributes to the recursion kernel, we take
\begin{equation}
	y(z)=\sum_{k=0}^{\infty}u_{2k-1}z^{2k-1}.
\end{equation}
Together with the standard Bergman kernel \eqref{Bergman kernel}, this gives
\begin{equation}
	\mathbb{K}(z_0,z)
	=\frac{\d z_0}{\d z}\,
	\frac{1}{\displaystyle\sum_{k=0}^{\infty}u_{2k-1}z^{2k-1}}\,
	\frac{1}{z_0^2-z^2}.
\end{equation}
Requiring
\begin{equation}
	\mathbb{K}(z_0,z)=K(z_0,z;\mathbf{s})\frac{\d z_0}{\d z}
\end{equation}
therefore gives
\begin{equation}\label{yfunction super high}
	\begin{aligned}
		\sum_{k=0}^{\infty}u_{2k-1}z^{2k}
		&=\left(
		\sum_{\mathbf{L}\in\mathbb{N}^\infty}
		z^{2|\mathbf{L}|}\beta_\mathbf{L}\mathbf{s}^\mathbf{L}
		\right)^{-1}\\
		&=\sum_{\mathbf{L}\in\mathbb{N}^\infty}
		\frac{(-1)^{\|\mathbf{L}\|}}{(2|\mathbf{L}|-1)!!}
		\frac{\mathbf{s}^{\mathbf{L}}}{\mathbf{L}!}
		z^{2|\mathbf{L}|},
	\end{aligned}
\end{equation} 
where the second equality follows from \eqref{beta inverse}.

Equation \eqref{yfunction super high} proves Theorem~\ref{spectral curve for hsv}. Moreover, the Eynard--Orantin multidifferentials are related to the generating functions by
\begin{equation}
	\omega^\Theta_{g,n}(z_1,\ldots,z_n;\mathbf{s})
	=W^\Theta_{g,n}(z_1,\ldots,z_n;\mathbf{s})
	\prod_{i=1}^n\d(-z_i).
\end{equation}

In particular, $u_{-1}=1$. Setting $s_i=0$ for every $i\geq 1$ reduces the spectral curve to the Bessel curve \eqref{Bessel curve}, which satisfies $xy^2=\frac{1}{2}$. Thus we recover the result of Do and Norbury \cite{doTopologicalRecursionBessel2018}.

Setting $s_i=0$ for $i\geq 2$ gives the one-parameter family
\[
x(z)=\frac12z^2,\qquad
y(z)=\frac{\cos(\sqrt{2s_1}\,z)}{z}.
\]
Specializing further to $s_1=2\pi^2$, we recover Norbury's
spectral curve for the super Weil--Petersson volumes
\cite[Theorem~6.8]{norburyEnumerativeGeometryModuli2026}.
\begin{corollary}
	After the specialization $s_1=2\pi^2$, the spectral curve
	\[
	x(z)=\frac12z^2,\qquad
	y(z)=\frac{\cos(2\pi z)}{z},\qquad
	B(z_1,z_2)=
	\frac{\d z_1\,\d z_2}{(z_1-z_2)^2}
	\]
	generates the correlators associated with the super
	Weil--Petersson volume polynomials.
\end{corollary}

Finally, the Eynard--Orantin topological recursion defines the free energies $F_g$. For genus $g\geq 2$, they are defined by
\begin{equation}
	F_g:=\frac{1}{2-2g}\operatorname*{Res}_{z=0}
	\left(\int_0^z\omega_{0,1}\right)\omega_{g,1}(z).
\end{equation}

For the spectral curve obtained above,
\begin{equation}
	\begin{aligned}
		\int_{0}^{z}\omega_{0,1}
		&=\int_0^z
		\sum_{\mathbf{L}\in\mathbb{N}^\infty}
		\frac{(-1)^{\|\mathbf{L}\|}}{(2|\mathbf{L}|-1)!!}
		\frac{\mathbf{s}^{\mathbf{L}}}{\mathbf{L}!}
		\xi^{2|\mathbf{L}|}\d\xi\\
		&=\sum_{\mathbf{L}\in\mathbb{N}^\infty}
		\frac{(-1)^{\|\mathbf{L}\|}}{(2|\mathbf{L}|+1)!!}
		\frac{\mathbf{s}^{\mathbf{L}}}{\mathbf{L}!}
		z^{2|\mathbf{L}|+1}.
	\end{aligned}
\end{equation}
Using the relation between $\omega^\Theta_{g,1}$ and $W^\Theta_{g,1}$, we obtain
\begin{equation}
	\begin{aligned}
		F_g(\mathbf{s})
		&=\frac{1}{2g-2}\operatorname*{Res}_{z=0}
		\Bigg(
		\sum_{\mathbf{L}\in\mathbb{N}^\infty}
		\frac{(-1)^{\|\mathbf{L}\|}}{(2|\mathbf{L}|+1)!!}
		\frac{\mathbf{s}^{\mathbf{L}}}{\mathbf{L}!}
		z^{2|\mathbf{L}|+1}\\
		&\hspace{80pt}\times
		\sum_{a\geq 0}\sum_{\mathbf{L}'\in\mathbb{N}^\infty}
		\left\langle\kappa(\mathbf{L}')\tau_a\right\rangle_g^\Theta
		\frac{\mathbf{s}^{\mathbf{L}'}}{\mathbf{L}'!}
		\frac{(2a+1)!!}{z^{2a+2}}
		\Bigg)\\
		&=\frac{1}{2g-2}
		\sum_{\mathbf{b}\in\mathbb{N}^\infty}
		\frac{\mathbf{s}^\mathbf{b}}{\mathbf{b}!}
		\sum_{\mathbf{L}+\mathbf{L}'=\mathbf{b}}
		(-1)^{\|\mathbf{L}\|}
		\binom{\mathbf{b}}{\mathbf{L}}
		\left\langle\tau_{|\mathbf{L}|}\kappa(\mathbf{L}')\right\rangle_g^\Theta\\
		&=\sum_{\mathbf{b}\in\mathbb{N}^\infty}
		\left\langle\kappa(\mathbf{b})\right\rangle_g^\Theta
		\frac{\mathbf{s}^\mathbf{b}}{\mathbf{b}!}.
	\end{aligned}
\end{equation}
In the last equality, we used the identity
\begin{equation}
	\left\langle\kappa(\mathbf{b})\right\rangle_g^\Theta
	=\frac{1}{2g-2}
	\sum_{\mathbf{L}+\mathbf{L}'=\mathbf{b}}
	(-1)^{\|\mathbf{L}\|}
	\binom{\mathbf{b}}{\mathbf{L}}
	\left\langle\tau_{|\mathbf{L}|}\kappa(\mathbf{L}')\right\rangle_g^\Theta,
\end{equation}
proved in \cite[Proposition~4.2]{huangHigherWeilPeterssonVolumes2025}. Hence, for genus $g\geq 2$, the Eynard--Orantin free energies coincide with the generating functions of the super higher Weil--Petersson volumes without $\psi$-class insertions. However, the original definitions of the genus-zero and genus-one free energies are not directly applicable to the local deformed spectral curve considered here.


\begin{thebibliography}{99}
	
	\bibitem{ahmedDiscreteAnalogue$mathscrN2$2026}
	W. Ahmed and J. Naskar.
	\emph{A Discrete Analogue of $\mathscr{N}=2$
	Super Weil--Petersson Volumes}. 2026. arXiv:
	\href{https://arxiv.org/abs/2610.13667}
	{\nolinkurl{2610.13667} \texttt{[hep-th]}}.
	

	\bibitem{arbarelloCombinatorialAlgebrogeometricCohomology1994}
	E. Arbarello and M. Cornalba. “Combinatorial and Algebro-Geometric
	Cohomology Classes on the Moduli Spaces of Curves”. In: \emph{Journal of
		Algebraic Geometry} 5.4 (1996), pp. 705–749.

	\bibitem{brezinExternalFieldProblem1980}
	É. Brézin and D. J. Gross. “The External Field Problem in the Large {$N$} Limit
	of {QCD}”. In: \emph{Physics Letters B} 97.1 (1980), pp. 120–124.

	
	\bibitem{chekhovTopologicalRecursionHard2019}
L. Chekhov and P. Norbury.
``Topological Recursion with Hard Edges''.
In: \emph{International Journal of Mathematics}
30.3, 1950014(2019).
	

	\bibitem{chidambaramRelations$overlinemathcalM_gn$Negative2025}
	N. K. Chidambaram, E. Garcia-Failde, and A. Giacchetto. “Relations on
	$\overline {\mathcal {M}}_{g,n}$ and the Negative {$r$}-Spin Witten Conjecture”.
	In: \emph{Inventiones Mathematicae} 241.3 (2025), pp. 929–997.

	\bibitem{dijkgraafLoopEquationsVirasoro1991}
	R. Dijkgraaf, H. Verlinde, and E. Verlinde. “Loop Equations and {Virasoro}
	Constraints in Non-Perturbative Two-Dimensional Quantum Gravity”. In:
	\emph{Nuclear Physics B} 348.3 (1991), pp. 435–456.

	\bibitem{doTopologicalRecursionBessel2018}
	N. Do and P. Norbury. “Topological Recursion on the {Bessel} Curve”. In:
	\emph{Communications in Number Theory and Physics} 12.1 (2018), pp. 53–73.

	\bibitem{eynardIntersectionNumbersSpectral2011}
	B. Eynard. \emph{Intersection Numbers of Spectral Curves}. 2011. arXiv: \href
	{https://arxiv.org/abs/1104.0176} {\nolinkurl {1104.0176} \texttt{[math-ph]}}.

	\bibitem{eynardInvariantsSpectralCurves2014}
	B. Eynard. “Invariants of Spectral Curves and Intersection Theory of Moduli
	Spaces of Complex Curves”. In: \emph{Communications in Number Theory and
		Physics} 8.3 (2014), pp. 541–588.

	\bibitem{eynardInvariantsAlgebraicCurves2007}
	B. Eynard and N. Orantin. “Invariants of Algebraic Curves and Topological
	Expansion”. In: \emph{Communications in Number Theory and Physics} 1.2 (2007),
	pp. 347–452.

	\bibitem{faberAlgorithmsComputingIntersection1999}
	C. Faber. “Algorithms for Computing Intersection Numbers on Moduli Spaces of
	Curves, with an Application to the Class of the Locus of Jacobians”. In: \emph{New
		Trends in Algebraic Geometry}. Ed. by K. Hulek, M. Reid, C. A. M. Peters, and
	F. Catanese. London Mathematical Society Lecture Note Series 264. Cambridge:
	Cambridge University Press, 1999, pp. 93–110.

	\bibitem{kobeuniversityjapanGeneralizationsMirzakhanisRecursion2024}
	H. Fuji and M. Manabe. “Some Generalizations of {Mirzakhani}’s Recursion and
	{Masur–Veech} Volumes via Topological Recursions”. In: \emph{Symmetry,
		Integrability and Geometry: Methods and Applications} 20, 043 (2024).

	\bibitem{grossPossibleThirdorderPhase1980}
	D. J. Gross and E. Witten. “Possible Third-Order Phase Transition in the
	Large-{$N$} Lattice Gauge Theory”. In: \emph{Physical Review D} 21.2 (1980),
	pp. 446–453.

	\bibitem{huangHigherWeilPeterssonVolumes2025}
	X. Huang, K. Liu, and H. Xu. \emph{Higher {Weil–Petersson} Volumes of the
		Moduli Space of Super Riemann Surfaces}. 2025. arXiv: \href
	{https://arxiv.org/abs/2501.05806} {\nolinkurl {2501.05806} \texttt{[math.AG]}}.

	\bibitem{jackiwLowerDimensionalGravity1985}
R. Jackiw. ``Lower Dimensional Gravity''.
In: \emph{Nuclear Physics B}
252 (1985), pp. 343--356.

	\bibitem{kaufmannHigherWeilPeterssonVolumes1996}
	R. M. Kaufmann, Y. I. Manin, and D. Zagier. “Higher {Weil–Petersson} Volumes
	of Moduli Spaces of Stable {$n$}-Pointed Curves”. In: \emph{Communications in
		Mathematical Physics} 181.3 (1996), pp. 763–787.

	\bibitem{kontsevichIntersectionTheoryModuli1992}
	M. Kontsevich. “Intersection Theory on the Moduli Space of Curves and the
	Matrix {Airy} Function”. In: \emph{Communications in Mathematical Physics}
	147.1 (1992), pp. 1–23.

	\bibitem{liuRecursionFormulaeHigher2008}
	K. Liu and H. Xu. “Recursion Formulae of Higher {Weil–Petersson} Volumes”. In:
	\emph{International Mathematics Research Notices} 2009.5 (2009), pp. 835–859.

	\bibitem{mirzakhaniSimpleGeodesicsWeilPetersson2006}
	M. Mirzakhani. “Simple Geodesics and {Weil–Petersson} Volumes of Moduli
	Spaces of Bordered Riemann Surfaces”. In: \emph{Inventiones Mathematicae} 167.1
	(2007), pp. 179–222.

	\bibitem{mirzakhaniWeilPeterssonVolumesIntersection2006}
	M. Mirzakhani. “{Weil–Petersson} Volumes and Intersection Theory on the
	Moduli Space of Curves”. In: \emph{Journal of the American Mathematical
		Society} 20.1 (2007), pp. 1–23.

	\bibitem{mulaseMirzakhanisRecursionRelations2006}
	M. Mulase and B. Safnuk. “{Mirzakhani}’s Recursion Relations, {Virasoro}
	Constraints and the {KdV} Hierarchy”. In: \emph{Indian Journal of Mathematics}
	50.1 (2008), pp. 189–218.

	\bibitem{norburyNewCohomologyClass2023}
	P. Norbury. “A New Cohomology Class on the Moduli Space of Curves”. In:
	\emph{Geometry \& Topology} 27.7 (2023), pp. 2695–2761.

	\bibitem{norburyEnumerativeGeometryModuli2026}
	P. Norbury. “Enumerative Geometry via the Moduli Space of Super Riemann
	Surfaces”. In: \emph{Journal of Geometry and Physics} 222, 105750 (2026).

	\bibitem{saadJTGravityMatrixIntegral2019}
	P. Saad, S. H. Shenker, and D. Stanford.
	\emph{JT Gravity as a Matrix Integral}. 2019.
	arXiv:
	\href{https://arxiv.org/abs/1903.11115}
	{\nolinkurl{1903.11115} \texttt{[hep-th]}}.

	
	\bibitem{stanfordJTGravityEnsembles2020}
	D. Stanford and E. Witten.
	``JT Gravity and the Ensembles of Random Matrix Theory''.
	In: \emph{Advances in Theoretical and Mathematical Physics} 24.6 (2020), pp. 1475--1680.

	\bibitem{teitelboimGravitationHamiltonianStructure1983}
	C. Teitelboim.
	``Gravitation and Hamiltonian Structure in Two Spacetime Dimensions''.
	In: \emph{Physics Letters B} 126.1--2 (1983), pp. 41--45.



	\bibitem{wittenTwodimensionalGravityIntersection1990}
	E. Witten. “Two-Dimensional Gravity and Intersection Theory on Moduli Space”.
	In: \emph{Surveys in Differential Geometry} 1.1 (1991), pp. 243–310.

	\bibitem{xiongEynardOrantinTopologicalRecursion2026}
	D. Xiong and J. Zhou. \emph{Eynard--{{Orantin Topological Recursion}}, {{Moduli-Space Virasoro Constraints}}, and {{Semisimple CohFTs}}}. 2026. arXiv: \href
	{https://arxiv.org/abs/2610.24980} {\nolinkurl {2610.24980} \texttt{[math-ph]}}. 
	

	\bibitem{zhouTopologicalRecursionsEynard2013}
	J. Zhou. “Topological Recursions of {Eynard–Orantin} Type for Intersection
	Numbers on Moduli Spaces of Curves”. In: \emph{Letters in Mathematical Physics}
	103.11 (2013), pp. 1191–1206.

	\bibitem{zhouTopologicalRecursionVirasoro2024}
	J. Zhou. \emph{Topological {{Recursion}} and {{Virasoro Constraints}} for {{Intersection Numbers}}}. Preprint.
  


\end{thebibliography}
\end{document}